\documentclass{emulateapj}
\usepackage{apjfonts}
\usepackage{amsmath}
\usepackage{graphicx}
\usepackage{dcolumn}

\newcommand{\pd}{\partial}  
\newcounter{ichi}
\newcounter{ni}
\newcounter{san}
\newcounter{yon}
\makeatletter
\newsavebox{\@parc@ption}
\def\parcaption#1{%
\sbox{\@parc@ption}{\shortstack[l]{#1}}%
>\setbox\@tempboxa\hbox{\csname fnum@\@captype\endcsname}%
\@tempdima\columnwidth \advance\@tempdima-\wd\@tempboxa
\@tempdimb\@tempdima 
\ifdim\wd\@parc@ption>\@tempdimb \@tempdima\@tempdimb
\else\@tempdima\wd\@parc@ption\fi
\sbox{\@tempboxa}{\parbox[t]{\@tempdima}{#1}}%
\caption{\usebox{\@tempboxa}}}
\makeatother

\slugcomment{}

\shorttitle{
The Role of Stochastic Acceleration in the GRB Prompt Emission
}
\shortauthors{Murase et al.}

\begin{document}
\title{
The Role of Stochastic Acceleration in the Prompt Emission of Gamma-Ray Bursts: 
Application to Hadronic Injection 
}


\author{
Kohta Murase\altaffilmark{1,2},
Katsuaki Asano\altaffilmark{2}, 
Toshio Terasawa\altaffilmark{2,3},
and Peter M\'esz\'aros\altaffilmark{4}
 }


\altaffiltext{1}{Dept. of Physics; Center for Cosmology and AstroParticle Physics, Ohio State University, 191 West Woodruff Avenue, Columbus, OH 43210, USA}
\altaffiltext{2}{Dept. of Physics, Tokyo Institute of Technology, 2-12-1 Ookayama, Meguro-ku, Tokyo 152-8550, Japan}
\altaffiltext{3}{Institute for Cosmic Ray Research, University of Tokyo, Kashiwa-no-ha 5-1-5, Kashiwa-shi, Chiba 277-8582, Japan}
\altaffiltext{4}{Dept. of Astronomy \& Astrophysics; Dept. of Physics; Center for Particle Astrophysics, Pennsylvania State University, University Park, PA 16802, USA}

\begin{abstract}
We study effects of particle re-acceleration (or heating) in the post-shock region via magnetohydrodynamic/plasma turbulence, in the context of a mixed hadronic-leptonic model for the prompt emission of gamma-ray bursts (GRBs), using both analytical and numerical methods.  We show that stochastically accelerated (or heated) leptons, which are injected via $pp$ and $p \gamma$ reactions and subsequent pair cascades, are plausibly able to reproduce the Band function spectra with $\alpha \sim 1$ and $\beta \sim 2-3$ in the $\sim$~MeV range.  An additional hard component coming from the proton-induced cascade emission is simultaneously expected, which is compatible with observed extra power-law spectra far above the MeV range.  We also discuss the specific implications of hadronic models for ongoing high-energy neutrino observations.
\end{abstract}


\section{Introduction} 
The origin of prompt emission from gamma-ray bursts (GRBs) has been an open issue for more 
than 40 years.  It is mainly observed in the $E\sim E^{\rm br} \sim 10~{\rm keV}-1~
{\rm MeV}$ range, most spectra being fitted by a smoothed broken power-law (the so-called 
Band function) with break energy $E^{\rm br}$.  The typical low-energy photon index is 
$\alpha \sim 1$ (where $d F/d E \propto E^{-\alpha}$ below $\sim E^{\rm br}$), while the typical high-energy photon index is $\beta \sim 2-3$ (where $d F/d E \propto E^{-\beta}$ above $\sim E^{\rm br}$).  Observed light curves are highly variable.  The variability can sometimes be $\sim$~ms, but pulses with various widths are observed.  For typical long GRBs, light curves consist of many variable pulses, so that the duration, $T \sim {10}^{1-2}$~s, is usually longer.  So far, many models have been proposed to explain this most luminous phenomenon in our universe~\citep[see recent reviews, e.g.,][]{Mes06,Zha07}. 
     
The classical scenario of the prompt emission is the optically-thin synchrotron internal shock model~\citep[e.g.,][]{RM94}, where shocks are responsible for dissipation of the outflow kinetic energy and observed gamma rays are attributed to synchrotron radiation from relativistic electrons.  
However, there are a couple of issues in the classical scenario.  First, the low-energy photon index may be incompatible with values predicted by synchrotron emission~\cite[e.g.,][]{GC99,MR00}.  This is the case especially when relativistic electrons are in the fast-cooling regime, which leads to $\alpha \sim 1.5$, although the Klein-Nishina effect on synchrotron self-Compton (SSC) process may reproduce $\alpha \sim 1$~\citep[e.g.,][]{DKK01,WLDM09,DBD11}.  One of the possible resolutions of this fast-cooling problem is to consider the magnetic field decay timescale that is comparable to the electron cooling timescale~\cite{PZ06}.  Injection of electrons to acceleration processes is one of the related issues.  Shock acceleration of primary electrons may be inefficient for sufficiently small emission radii because of the rapid cooling in the strong magnetic field region at the vicinity of the shock~\cite{MS09}.  If the outflow is magnetized enough, even for sub-luminal shocks, recent \textit{ab intio} particle-in-cell simulations of collisionless shocks indicate that acceleration of nonthermal electrons at relativistic shocks are inefficient whereas ions can be accelerated~\cite{SS11}.    
Another issue is the radiative efficiency problem~\citep[e.g.,][]{ITYN06,Zha+07}, 
although this depends on estimates of the afterglow kinetic energy~\cite{EW05}.  The high efficiency can be achieved by a large dispersion in the Lorentz factor of the outflows~\citep[e.g.,][]{Bel00}, but it does not seem so easy to reconcile that with the observed spectral correlations~\citep[e.g.,][]{Yon+04}.  A recent discussion of the classical scenario is found in Daigne et al. (2011). 
     
On the other hand, many alternative models have been suggested.  Diffusive synchrotron radiation can explain the low-energy photon index if the fast-cooling problem is fixed~\cite{Med00}.  Instead of dissipation of the kinetic energy via internal shocks, dissipation of the magnetic energy via, e.g., magnetic reconnection may play a crucial role as the dissipation mechanism, and observed gamma rays may be synchrotron or other electromagnetic radiation from relativistic electrons~\citep[e.g.,][]{SDD01,Lyu06,ZY11,MU11}.   
If significant thermal energy is stored in the GRB jet by energy deposition from the central engine and/or via some dissipation below the photosphere, the prompt emission may originate from quasi-thermal emission caused by Comptonization or Coulomb heating~\citep[e.g.,][]{Tho94,GC99,MR00,PMR06,Iok+07,Bel10}.  
One of the appealing possibilities is to consider the slow heating scenario.  Around the photosphere, electrons that are slowly heated can lead to $\alpha \sim 1$ via Comptonization~\cite{Tho94,PMR06}, where the typical electron energy is determined by the balance between heating and cooling~\cite{GC99,VP09}.  Alternatively, in the neutron-rich outflow, Coulomb collisions should work as slow heating below the baryonic photosphere, leading to hard photon spectra of $\alpha \sim -0.5-1$ and $\beta \sim 2.5$ via the pair injection by the $np$ reaction and subsequent cascades~\cite{Bel10,VBP11}.  Such slow heating may be caused by stochastic acceleration (so-called second-order Fermi acceleration) over the sub-hydrodynamical timescale and/or hydrodynamical timescale, which may occur in the magnetohydrodynamic (MHD) and plasma turbulences generated by shock or magnetic dissipation such as magnetic reconnections~\cite{BM96,AT09,ZY11}.  

Emission at higher energies above $\sim 10$~MeV should also be important to reveal the radiation mechanism of the prompt emission, and various high-energy processes such as Comptonized thermal, SSC~\cite{RM94,AI07}, proton-induced cascade~\cite{Vie97,DA06}, and proton synchrotron~\cite{Tot98} emissions, were discussed.  Such high-energy emission had been sparsely detected by the EGRET onboard the Compton Gamma-Ray Observatory~\citep[e.g.,][]{Hur+94}.  Recently, however, significant observational progress has been achieved by \textit{Fermi}.  Dozens of GRBs have been observed by LAT onboard \textit{Fermi}, and some bursts can be fitted by the Band function up to GeV energies~\cite{Abd+09a,Abd+09c,Abd+10}, while some of them (GRB 090510, 090902B, 090926A) clearly have an extra hard component at $\gtrsim 10$~MeV~\cite{Abd+09b,Ack+10,Ack+11}.  GRBs with the extra GeV component belong to the brightest class of GRBs, which may indicate that this could be more common, even though it can be seen only in bright LAT GRBs~\cite{Gra10}.  
Another feature found by \textit{Fermi} is that $\gtrsim 100$~MeV gamma rays are delayed by $\sim 0.1-1$~s in the cosmological rest frame, behind the onset of the MeV emission.   

The origin of the high-energy emission is now under active debate.  
Late-time high-energy gamma-ray emission from GRBs such as 080916C, 090510, and 090902B has been attributed to afterglow emission rather than the prompt emission~\cite{KD10,Ghi+10,Raz11}.  But, the high-energy emission in the early phase usually has a strong variability and often correlates with the MeV emission, which indicates an internal origin~\cite{He+11}.  Since the simple one-zone leptonic model often has difficulties, multi-zone models, not only the SSC model~\cite{CGP10,DBD11} but also the external inverse-Compton (IC) model~\cite{TWM11,Pee+11} and the synchrotron model~\cite{Iok10}, have been invoked.  The one-zone hadronic models are also viable, and the high-energy emission can be explained by proton-induced cascade or proton synchrotron emission~\cite{AGM09,RDF10}.  The hadronic models are of interest, since GRBs may be the sources of observed ultra-high-energy cosmic rays (UHECRs), whose origin has been a big mystery for about 50 years~\cite{Wax95,Vie95,MINN08}, and their neutrino signals may be seen by ongoing neutrino observations~\citep[a recent review is][]{Wax11}.

In this paper, we study a slow heating scenario in the presence of electromagnetic cascades initiated by injection at high energies.  To explain the MeV emission self-consistenly, we consider effects of acceleration (via the second-order Fermi mechanism) or heating of secondaries, which can be anticipated in the post-shock region of relativistic shocks (or magnetic reconnection).  We show that cascades can play an important role in the injection of re-accelerated/heated particles, and slow heating scenarios via stochastic acceleration or wave heating typically lead to hard spectra compatible with observations.  As an injection process at high energies, we consider an application to hadronic injection via $pp$ and $p \gamma$ reactions, whereby the hadronic models have the appealing feature of explaining an extra hard component at GeV energies.   

The paper is organized as follows.  Section~2 describes an overview of the model, and we provide the numerical method and its results in Section~3.  We discuss implications for neutrinos in Sections~4, and our results are summarized in Section~6.  Throughout this work, cosmological parameters are set to $H_0=71~{\rm km}~{\rm s}^{-1}~{\rm Mpc}^{-1}$, $\Omega_M=0.3$, and $\Omega_\Lambda=0.7$, and we adopt the conventional notation $Q=Q_x \times {10}^{x}$.  

\section{Theoretical Model}
In this work, we investigate the role of particle re-acceleration, motivated by the goal to resolve the low-energy photon index and fast-cooling problems.  In the post-shock region of relativistic shocks, turbulent magnetic fields would be generated, where stochastic acceleration (or second-Fermi acceleration) may work efficiently since the Alfv\'en velocity is expected to be close to $\sim (0.1-1) c$.  The second-order Fermi acceleration over the sub-hydrodynamical scale plays a role of slow heating to avoid the fast cooling problem of relativistic electrons, and it may give harder electron spectra than the typical first-order Fermi acceleration. This possibility was proposed by Bykov \& M\'esz\'aros (1996) and recently studied by Asano \& Terasawa (2010).  Also, particles may be slowly heated by MHD waves, as has been discussed in the solar corona~\citep[e.g.,][]{Alf47,FC05,SI05,Tom+07,Jes+09} 

Such slow heating operates not only for primary electrons but also electron-positron pairs generated via cascades initiated by some high-energy injection.  Motivated by the electron injection/acceleration problem and the existence of an extra hard component at $\gtrsim$~10 MeV, this work focuses on a hadronic model, where pairs are produced via the electromagnetic cascade initiated by hadronic injection from $pp(np)$ and $p \gamma$ reactions, i.e., the proton-induced cascade.  Although there is a similarity to the collisional scenario proposed by Beloborodov (2010), our scenarios are different in that leptons are heated by turbulence rather than Coulomb heating and the dissipation radius can be much larger. 

The hadronic emission associated with prompt emission was originally motivated by the hypothesis that observed UHECRs come from GRBs~\cite{Wax95,Vie95,MINN08}.  This requires that the UHECR energy (at $\sim {10}^{19}$~eV) is comparable to the gamma-ray energy, $\tilde{\mathcal E}_{\rm HECR}\equiv {E_{p}^{'}}^2 (dN_p/d E_p^{'}) \sim \mathcal{E}_{\gamma}$.  Although the GRB rate evolution, total gamma-ray energy and proton spectral index are still uncertain, the total nonthermal baryon loading is expected to be large, $\xi_{\rm CR} \equiv {\mathcal E}_{\rm CR}/{\mathcal E}_{\gamma} \sim 3-300$~\citep[e.g.,][and references therein]{MINN08,Wax10}.  If sufficiently large baryon loading is realized and the photomeson production efficiency is high enough, the hadronic emission can dominate over the leptonic emission at high energies~\citep{AIM09}, and the visibility of the hadronic components at high energies is enhanced when the high-energy photon spectrum is much steeper than $\beta \sim 2$.  The extra hard component in the GeV range, which was seen by \textit{Fermi}, can be explained by the hadronic emission~\cite{AGM09}.  But, in the previous work, the Band function was assumed \textit{ad hoc}, otherwise the low-energy photon index becomes too soft (which leads to overestimating gamma rays and neutrinos).  

In the next subsection, we consider shock dissipation in the baryonic outflow and shock acceleration of particles.  Then, we discuss stochastic acceleration (or heating) by plasma/MHD turbulence in the downstream region to take into account its effects with a simplified model.  

\subsection{Shock Acceleration at Internal Shocks}
In the classical scenario, electrons are accelerated at internal shocks, and the observed gamma-ray emission is attributed to electromagnetic radiation from non-thermal electrons. If the outflow contains baryons, it is natural to expect that protons are also accelerated on a timescale
\begin{equation}
t_{\rm acc} = \eta \frac{r_L}{c},
\end{equation}
where $\eta \sim 1-10$ in the efficient case~\citep[e.g.,][]{BO96,RM98}.  The acceleration timescale is typically fast, and protons that are accelerated up to sufficiently high energies should interact with non-thermal photons via photomeson production, leading to $\sim \rm PeV$ neutrinos~\cite{WB97,RM98}, GeV-TeV gamma rays~\cite{Vie97}, and UHE gamma rays~\cite{Mur09}. Around the photosphere, GeV-TeV neutrinos and GeV-TeV gamma rays are also produced through the $pp(np)$ reaction~\cite{Mur08,WD09}. 

In our model, we consider a similar picture in which charged particles are accelerated at shocks.  Motivated by the electron injection/acceleration issue, however, we focus on the case of hadronic injection (neglecting primary electrons), where proton-induced gamma-ray emission is especially relevant.  When the shock dissipation radius is small enough, the efficient $pp(np)$ reaction occurs in the baryon-rich outflow, and high-energy gamma rays and pairs are produced as well as neutrinos. Also, sufficiently high-energy protons can interact with soft photons which may come from residual thermal emission around the photosphere~\cite{AT03} (and/or non-thermal emission from primary electrons if their acceleration occurs).  Those $pp$ and $p \gamma$ reactions lead to production of high-energy gamma rays with $E \sim 0.1 E_p \simeq {\rm TeV}~ E_{p,12}$ and pairs with $E_e \sim 0.05 E_p \simeq  0.5~{\rm TeV}~ E_{p,12}$.  Lower-energy photons should also be produced via the synchrotron and IC processes, which eventually affect the distribution of seed photons so that the processes are nonlinear.  
  
Sufficiently high-energy gamma rays should lead to pair creation with seed photons at $\tilde{E} E \approx \Gamma^2 m_e^2 c^4$, so that the following cascade is unavoidable.  As long as the pair-creation opacity of high-energy gamma rays is larger than unity, gamma rays lead to generation of pairs that lose their energies via synchrotron and IC emissions.  As a result, the cascade gamma-ray spectrum will be eventually formed, where the number of pairs can be larger than that of pairs injected directly from the $pp$ and $p \gamma$ reactions.  

We are interested in cases where the $pp$ or $p \gamma$ reaction efficiency is high enough, otherwise the hadronic component is insufficient to explain the high-energy gamma-ray emission observed by \textit{Fermi}.  Hence, for the purpose of demonstrations, this work focuses on relatively small dissipation radii of $r \sim {10}^{12-13}$~cm such that $\tau_T \approx \sigma_T n_{\rm th} l \sim 0.1-1$ (where $n_{\rm th}$ is the thermal proton density and $l$ is the comoving length), though large uncertainty allows us to consider different cases where the $pp$ and $p \gamma$ reaction efficiencies are much lower~\citep[e.g.,][]{MINN08}. 
The $pp$ reaction efficiency is roughly estimated to be $f_{pp} \approx \kappa_{pp} \sigma_{pp} n_{\rm th} l \simeq 0.05 \tau_T$~\cite{Mur08}, where $\kappa_{pp}$ is the $pp$ inelasticity of protons.  We expect that the observed gamma-ray spectrum will be eventually formed (see the next section). Then, assuming the (broken) power-law photon spectrum, the effective photomeson production efficiency is estimated to be~\cite{WB97,MN06b} 
\begin{equation}
f_{p \gamma} (E_p) \sim 0.7 \frac{L_{\gamma,51.5}^{\rm br}}{r_{13} \Gamma_{2.7}^2 (E^{\rm br}/500~\rm keV)} 
\left\{ \begin{array}{ll}
{(E_p/E_p^{\rm br})}^{\beta-1}
& \mbox{($E_p < E_p^{\rm br}$)}\\
{(E_p/E_p^{\rm br})}^{\alpha-1} 
& \mbox{($E_p^{\rm br} < E_p$)}.
\end{array} \right.
\end{equation}
where $E_p^{\rm br} \simeq 80~{\rm PeV}~{(E^{\rm br}/500~\rm keV)}^{-1} \Gamma_{2.7}^2$ is the energy of protons interact with photons with $E^{\rm br}$ and the multi-pion production effect is included. 

Gamma rays also interact with the same target photon field, so that the above equation implies that the pair-creation opacity is~\citep[e.g.,][]{LS01,MI08,GZ08} 
\begin{eqnarray}
\tau_{\gamma \gamma} (E) &\approx& \frac{f(\beta) \sigma_T l L_{\gamma}^{\rm br}}{4 \pi r^2 \Gamma c E^{\rm br}} {\left( \frac{E E^{\rm br}}{\Gamma^2 m_e^2 c^4} \right)}^{\beta-1} \nonumber \\
&\sim& 280~\frac{L_{\gamma,51.5}^{\rm br}}{r_{13} \Gamma_{2.7}^2 (E^{\rm br}/500~\rm keV)} 
\left\{ \begin{array}{ll}
{(E/\tilde{E}^{\rm br})}^{\beta-1}
& \mbox{($E < \tilde{E}^{\rm br}$)}\\
{(E/\tilde{E}^{\rm br})}^{\alpha-1} 
& \mbox{($\tilde{E}^{\rm br} < E$)}.
\end{array} \right.
\end{eqnarray}
where $\tilde{E}^{\rm br} \simeq 130~{\rm GeV}~{(E^{\rm br}/500~\rm keV)}^{-1} \Gamma_{2.7}^2$ and $f (\beta) \sim 0.1$ for $\beta \sim 2$~\cite{Bar06}. (Note that the opacity in the highest energies can be smaller than unity due to existence of the synchrotron self-absorption cutoff.)  Then, in a simple one-zone model, the pair-creation break (or cutoff) is estimated to be $E^{\rm cut} \simeq 1.2~{\rm GeV}~{(L_{\gamma,51.5}^{\rm br})}^{-\frac{1}{\beta-1}} r_{13}^{\frac{1}{\beta-1}} \Gamma_{2.7}^{\frac{2 \beta}{\beta-1}} {(E^{\rm br}/500~\rm keV)}^{\frac{2-\beta}{\beta-1}}$.  Most of the high-energy gamma rays are radiated below this energy, and the above equation suggests that the effective injection rate of electron-positron pairs is 
\begin{equation}
\varepsilon Q_{i} \sim \frac{{\rm min}[1,<f_{\rm mes}>] U_p G_\varepsilon}{t_{\rm dyn}}, 
\end{equation}
where $U_p$ is the total cosmic-ray proton energy density, $t_{\rm dyn}\approx 3 l/c$ is the dynamical time, $<f_{\rm mes}>$ is the effective meson production efficiency averaged over proton energies, and $G_\varepsilon$ such that $\int d\varepsilon G_\varepsilon = 1$ is determined by the details of the pair cascade. Because of the cascade and the multiplicity of the hadronic reactions, the number of pairs can be even larger than that of the reactions~\citep[cf.][]{Bel10}.  If there is no re-acceleration/heating and escape, we get 
\begin{equation}
\frac{\pd n_{\varepsilon}}{\pd t} = \frac{\pd}{\pd \varepsilon} \left[ \frac{d \varepsilon}{dt} n_{\varepsilon} \right] +Q_{i},
\end{equation}
where
\begin{equation}
\frac{d \varepsilon}{dt} = - \frac{4}{3} \sigma_T c (U_B+U_{\rm KN}) (\gamma^2-1)  
\end{equation}
and $U_B$ is the magnetic field energy density and $U_{\rm KN}$ is the photon energy density with the correction by the Klein-Nishina effect. 
Therefore, at energies where injection occurs, we expect that the cascade typically leads to a flat energy spectrum of electrons, so that we obtain $E F(E) \propto E^{0-0.5}$ that is also seen by our numerical calculations.  Such hard photon spectra can explain the extra hard component at $\gtrsim 10$~MeV observed by \textit{Fermi}~\citep[e.g.,][]{AGM09}. 
In the hadronic model that attributes the extra high-energy component to the proton-induced cascade, one typically expects
\begin{equation}
{\mathcal E}_{\gamma \rm ex} \sim {\rm min}[1,<f_{\rm mes}>] {\mathcal E}_{\rm CR},   
\end{equation}
where $\mathcal{E}_{\gamma \rm ex}$ is the energy in gamma-rays of the extra component, which is typically smaller than the total gamma-ray energy released as prompt emission, ${\mathcal E}_{\gamma}$.  As the emission radius is larger (leading to lower $<f_{\rm mes}>$), larger ${\mathcal E}_{\rm CR}$ is required.  
 
\subsection{Slow Heating by Turbulence}
Next, we consider stochastic acceleration of particles that are produced by injection at high energies and subsequent cascades.  The turbulent magnetic field in the downstream of the shock is expected to be significantly amplified.  
Cascade pairs that are rapidly cooling (i.e., they are in the fast cooling regime) are distributed in the downstream region, and can be re-accelerated by plasma/MHD turbulence via the second-order Fermi mechanism.  Assuming isotropy of the particle momenta and of the fluctuations, one may apply the Fokker-Planck equation~\citep[e.g.,][]{LMPF06,BL07},
\begin{equation}
\frac{\pd n_{\varepsilon}}{\pd t} = \frac{\pd}{\pd \varepsilon} \left( D_{\varepsilon \varepsilon} \frac{\pd n_{\varepsilon}}{\pd \varepsilon} \right) - \frac{\pd}{\pd \varepsilon} \left[ \left(A-\frac{d \varepsilon}{dt} \right) n_{\varepsilon} \right] -\frac{n_{\varepsilon}}{t_{\rm esc}}+Q_{i},
\end{equation}
where $A=(D_{\varepsilon \varepsilon}/\varepsilon) (1-\gamma^{-2})/(1+\gamma^{-1})$ and $D_{\varepsilon \varepsilon}$ is the diffusion coefficient.  For isotropic magnetic field fluctuations with spectral energy density $W_k \propto k^{-\tilde{q}}$, one may write $D_{\varepsilon \varepsilon} \propto \varepsilon^{q}$ with $q=\tilde{q}$ for resonant acceleration, where scattering and acceleration timescales of charged particles are proportional to $\varepsilon^2/D_{\varepsilon \varepsilon} \propto \varepsilon^{2-q}$~\citep[e.g.,][]{MR89,DML96}.  For non-resonant acceleration, one may expect $D_{\varepsilon \varepsilon} \propto \varepsilon^{2}$~\cite{BL07}.
The fluid and particles behind the shock are advected to the downstream.
The length scale of the turbulent region will be $l_{\rm tur} \approx (c/3)t_{\rm tur}$, where $t_{\rm tur}$ is the typical lifetime of the turbulence.  Then, since the Alfv\'en velocity can be close to $\sim (0.1-1) c$ given that the magnetic field is significantly amplified ($\beta_A \equiv B/\sqrt{4 \pi h} \sim \sqrt{(3U_B/2U_{\rm th})} \simeq 0.4 \epsilon_{B,-1}$, where $h$ is the enthalpy density and $\epsilon_B$ is the energy fraction carried by magnetic fields), for fluctuations with a simple power law, the stochastic acceleration timescale is expressed as~\cite{DML96,BM96,PL04} 
\begin{equation}
t_{\rm sta} \equiv \frac{\varepsilon^2}{4 D_{\varepsilon \varepsilon}} = \eta_{\rm sta} \frac{l_{\rm tur}}{c} {\left( \frac{r_L}{l_{\rm tur}}\right)}^{2-q}, 
\end{equation}
where $r_L$ is the Larmor radius, $\eta_{\rm sta} \sim 1$ is a pre-factor that depends on the wave amplitude and the Alfv\'en velocity, and we have assumed that $2 \pi/k_{\rm min} \sim l_{\rm tur}$. 
This stochastic acceleration timescale can be longer than the shock acceleration timescale, $t_{\rm acc} = \eta r_L/c$.  But it can still be shorter than the hydrodynamical timescale, $l_{\rm tur}/c$, in the relevant energy range where $r_L \ll l_{\rm tur}$ (except for exactly $q=2$ over all $k$).  In this sense, the stochastic acceleration plays the role of a slow heating over the sub-hydrodynamical timescale~\cite{BM96}.  

Various numerical simulations, including plasma simulations for collisionless shocks~\citep[e.g.,][]{Spi08,CSA08,Kes+09} and MHD simulations~\citep[e.g.,][]{ZMW09,Miz+11,IAI11}, have recently been attempted.  
For the MHD scale turbulence, Inoue et al. (2011) showed that highly relativistic turbulence decays fast, whereas the transonic and subsonic turbulences are maintained much longer, and their timescales are determined by the initial scale of the inhomogeneity in the fluid density, which may be comparable to the shell width.
Such long-lasting large scale turbulences may play an important role to accelerate/heat high-energy electrons (and positrons) that emit MeV gamma-rays.

On the other hand, small scale turbulences induced via plasma instabilities, such as seen in PIC simulations~\citep[e.g.,][]{Spi08,CSA08} seem to settle quickly and exist only in the vicinity of the shock at earlier times.
Keshet et al. (2009) suggested high-energy particles play an important role for the long term evolution and the magnetic field is progressively generated on larger scales in a growing region around the shock, where collisionless shock configurations do reach the steady state.  
Effects of high-energy particles may be important for the generation and sustainment of magnetic field turbulences.  The growth of the turbulence scale, as seen in coalescence of current beams for the case of the Weibel instability (Kato 2005), should be tested through long-term simulations including the non-thermal particles.

Below, we mainly expect that the resonant acceleration via scatterings with Alfv\'en waves can work, whereas one might expect other possibilities such as the non-resonant acceleration or heating by plasma/MHD waves. 
However, the non-resonant acceleration due to compressive modes will not work for $\beta_A > \beta_{\rm cs}$ (sound speed divided by $c$) since the fast and slow modes will be damped due to the Landau damping, although one may have possibilities of some non-resonant scattering by Alfv\'en waves or magnetic bottles formed by large-scale slow-mode perturbations for $\beta_A < \beta_{\rm cs} \approx 1/\sqrt{3}$ (due to small values of $\epsilon_B$)~\cite{BYL11}~\footnote{The Landau damping rate would be small if the distribution function of thermal ions have negligible slopes, $\partial f/\partial v$, around the resonant velocity. But, in this case, the majority of ions could be trapped in finite amplitude slow-mode waves, so possible nonlinear effects should be incorporated to obtain the damping rate.}.
On the other hand, dissipation of MHD waves produced at the shock can play crucial roles on heating particles.  Heating of particles has been debated in the context of the solar corona~\citep[e.g.,][]{SI05,Tom+07,Jes+09} and advection-dominated accretion flow~\citep[e.g.,][]{Med00b}, but the details are uncertain in the case of the relativistic plasma expected for the GRB shock. Though the detailed study is beyond the scope of this work, we expect the analogous slow heating effect on forming a spectrum as long as particles in the relevant energy range can be heated by the waves.

At present, we still have no complete model of the turbulence for GRB internal shocks.
Hence, as in previous work~\cite{AT09}, we consider the simplest setup where the stochastic acceleration time is energy-independent ($D_{\varepsilon \varepsilon} \propto \varepsilon^{2}$) in the range in which we are interested.  Such a situation may be realized for the resonant acceleration, if $q \sim 1.5-2$ at relevant electron energies such that $r_L \ll l_{\rm tur}$, and necessary steep turbulent spectra can be expected for the MHD turbulence ($\tilde{q} \sim 1.5-2$)~\citep[e.g.,][]{Cho+03,LMPF06} or collisionless shocks in the presence of cosmic rays ($\tilde{q} \sim 2$)~\cite[e.g.,][]{KKW07}.

Note that the turbulent spectrum may not be a simple power law so $q$ in the diffusion coefficient is rather energy-dependent in the wide energy range, and the MHD simulations have suggested the spectrum is flat at large scales but steep enough at small scales above the transition scale~\citep[e.g.,][]{CSA08,IAI11}.  But it would be approximately valid to assume a singe value of $q$ in the relatively narrow energy range relevant for the MeV emission.  Although future numerical studies on plasma/MHD turbulence are required for more realistic calculations, our phenomenological approach is enough to demonstrate the role of stochastic acceleration and we show that the process efficiency is reasonably sufficient to explain observations given the setup.  

Next, we briefly discuss the expected spectra of particles re-accelerated via stochastic acceleration in the plasma/MHD turbulence, although the details depend on diffusion, energy losses, and particle escape.  Let us consider low energies where energy losses are negligible.  Then, for $q=2$, the steady-state spectrum is obtained as $n_\varepsilon \propto \varepsilon^{0.5-{(9/4+4t_{\rm sta}/t_{\rm esc})}^{1/2}}$~\cite{LMPF06}.  At sufficiently low energies ($r_L \ll l_{\rm tur}$) that are relevant in the case of GRB prompt emission, $t_{\rm esc}$ is so long that we may drop it, where one has $n_{\varepsilon} \propto \varepsilon^{1-q}$.  This means that the stochastic acceleration leads to hard electron spectra, and the synchrotron spectrum becomes  
\begin{equation}
E F(E) \propto E^{2-q/2} \,\,\, (E \leq E^{\rm br}), 
\end{equation}
for $q>4/3$ (and we get $E F(E) \propto E^{4/3}$ for $q \leq 4/3$).  Hence, the low-energy photon index of $\alpha \sim q/2 \sim 1$ is possible if $q \sim 1.5-2$ (which seems typical for the MHD turbulence).  In the classical scenario of prompt emission, electrons are typically in the fast cooling regime, so that high-energy electrons cool down within the advection time, $\sim 3l/c$.  But, this problem can be avoided when acceleration (or wave heating) plays a role of slow heating. 

When cooling is absent, the maximum energy of re-accelerated/heated electrons should eventually be limited, since they can carry only a fraction of the turbulent energy.  Or, radiative cooling is so strong that the typical electron energy is determined by the balance between acceleration/heating and cooling.  If we knew the turbulent spectrum from first principles, we would be able to calculate $\gamma_{\rm typ}$ for the stochastic acceleration.  For example, if the turbulent spectrum is flat at large scales and then becomes sufficiently steep at $\sim 100 k_{\rm min}$, the stochastic acceleration time can be $t_{\rm sta} \sim 1.3 \times {10}^{-5}~{\rm s}~(l_{\rm tur}/0.2 l) r_{13} \Gamma_{2.7}^{-1}$. Comparison with the synchrotron cooling time of $t_{\rm syn} \simeq 1.8 \times {10}^{-5}~{\rm s}~\epsilon_{B,-1}^{-1} r_{13}^2 \Gamma _{2.7}^2 L_{\rm th,53.5}^{-1} \gamma_{2.7}^{-1}$ ($L_{\rm th}$ is the luminosity associated with the thermal energy) gives $\gamma_{\rm typ} \sim 690~\epsilon_{B,-1}^{-1} {(l_{\rm tur}/0.2 l)}^{-1} r_{13} \Gamma_{2.7}^3 L_{\rm th,53.5}^{-1}$ leading to $\sim$~MeV synchrotron gamma rays.  
Unfortunately, however, the realistic stochastic acceleration rate is highly uncertain at present so that we instead make an \textit{ad hoc} assumption that the stochastic acceleration rate is determined by requiring that the observed break energy is the synchrotron energy of electrons with the typical electron Lorentz factor~\cite{AT09}, where we find\footnote{Phenomenologically, one can obtain another expression, $\gamma_{\rm typ} \sim {\left[ \frac{3-q}{2-q} \frac{\epsilon_e}{f_e} \frac{m_p}{m_e} \gamma_{p, \rm th} \gamma_{\rm reacc}^{1-q} \right]}^{\frac{1}{2-q}}$ for $q<2$ or $\gamma_{\rm typ} \sim \ln(\frac{\gamma_{\rm typ}}{\gamma_{\rm reacc}}) \frac{\epsilon_e}{f_e} \frac{m_p}{m_e} \gamma_{p, \rm th}$ for $q \sim 2$.
Here $\epsilon_e$ is the energy fraction carried by electrons accelerated via turbulence and $\gamma_{\rm reacc}$ is the electron Lorentz factor at which the stochastic acceleration does not operate (which can be expected when particles do not interact with plasma/MHD waves).  
Then, $f_e$ is the effective number fraction of electrons that are injected to the stochastic acceleration, which is determined by the details of the cascade and acceleration processes.} 
\begin{eqnarray}
\gamma_{\rm typ} &\approx& {\left( \frac{E^{\rm br} m_e c}{\Gamma \hbar e B} \right)}^{1/2} \nonumber \\ 
&\simeq& 550~r_{13}^{1/2} \epsilon_{B,-1}^{-1/4} {\left( \frac{E^{\rm br}}{500~\rm keV} \right)}^{1/2} L_{\rm th,53.5}^{-1/4}. 
\end{eqnarray}
As described in the next section, we take into account slow heating (stochastic acceleration or wave heating) of leptons by using the Monte Carlo method, assuming that the acceleration/heating time is energy-independent and equation~(11) applies.  Although it is a toy model, we can demonstrate that the re-accelerated/heated particles have hard spectra leading to $\alpha \sim 1$, as expected in this subsection.  

\section{Numerical Results of Gamma-Ray Spectra}
In the previous section, we gave the basic picture of our model with analytical expressions.  
But many microphysical processes have to be taken into account in order to calculate gamma-ray spectra.  One of the approaches is to solve the kinetic equations~\citep[e.g.,][]{Sve87}, but it is time consuming for our problem.  Alternatively, we employ the Monte Carlo code used in, e.g., Asano et al. (2009a), Asano \& Terasawa (2009), where the energy distributions of all particles are simulated iteratively until they converge to a self-consistent steady state, which is assumed to be realized within the pulse timescale. 

All the important microphysical processes for the electromagnetic cascade, $\gamma \gamma$ pair creation by electron and positron pairs, synchrotron, and Compton (both in the Thomson and Klein-Nishina regimes) emissions from all relativistic particles, and synchrotron self-absorption, are properly included~\citep[e.g.,][]{BG70}.  The photomeson production is approximately included as in Asano \& Nagataki (2006), using experimental data but neglecting high-multiplicity/inelasticity, multi-pion production at high energies, and it is assumed that a muon neutrino from pion decay carries a quarter of the pion energy, and neutrinos and a lepton from muon decay carry a third of of the muon energy.  Note that such a simplified treatment is timesaving and sufficient for this work, since effects of more detailed microphysics are typically moderate~\cite{MN06a,Mur07,BHW11}.  The Bethe-Heitler process by protons is also included using the cross section and inelasticity given by Chodorwski et al. (1992), but this effect is only modest in the relevant energy range for GRB prompt emission spectra~\citep[e.g.,][]{MPR00}.  In addition, we implement in detail the $pp$ reaction for which high multiplicity and inelasticity in pion production is important.  The high-energy process is calculated as in Murase (2008), whereas the low-energy process below $100$~GeV is included based on runs of Geant 4~\cite{Ago+03}.  The Thomson scattering effect due to thermal electrons, which would weaken photons of energy above $\sim m_{\rm e} c^2$ in the comoving frame, is included by assuming that photons interact with electrons of temperature $100$~keV within a timescale $l/c$ before they escape from the emission region.  The spectral softening by this effect has been calculated with a Monte-Carlo method with the Klein-Nishina cross section calculated in advance. 
 
There are several input parameters required for calculations. 
The emission radius $r$ and the bulk Lorentz factor $\Gamma$ are relevant parameters.  One typically expects $r \sim {10}^{13} - {10}^{15.5}$~cm in the classical internal shock scenario, or $r \sim {10}^{11}-{10}^{13}$~cm in the baryonic photospheric scenario.  We consider relatively small radii (but larger than the coasting radius), $r \sim {10}^{12}-{10}^{13}$~cm, where the efficient hadronic reactions are efficient~\cite{Asa05,MN06a}.  We also take $\Gamma \sim 100-1000$, and the comoving length is given by $l = r/\Gamma$.  
Another important parameter is the luminosity of protons accelerated at the shock, $L_p = 4 \pi r^2 \Gamma^2 c U_p$ (and the nonthermal baryon loading parameter is expressed as $\xi_{\rm CR} = L_p/L_\gamma$).  In this work, we consider $L_p \sim {10}^{53}-{10}^{54}~{\rm erg}~{\rm s}^{-1}$, assuming the proton distribution with $d n_p/d \varepsilon \propto \varepsilon_p^{-2} {\rm e}^{-(\varepsilon_p/\varepsilon_p^{\rm max})}$ for $\varepsilon_p \geq \gamma_{\rm th} m_p c^2 = 4 m_p c^2$.  The maximum energy $\varepsilon_p^{\rm max}$ is given by the comparison between the acceleration time and cooling timescales as well as the Hillas condition.  We also assume the luminosity associated with the thermal energy, $L_{\rm th}= 4 \pi r^2 \Gamma^2 c U_{\rm th} (\gtrsim L_p)$, where $U_{\rm th}=\gamma_{\rm th} n_{\rm th} m_p c^2$ is the thermal energy density.  The comoving magnetic field is introduced as $U_{B} (= \epsilon_B U_{\rm th}) \sim 0.01-1 U_p $.  In accord with our motivation and purposes, we neglect the acceleration of primary electrons.   
   
In Figure~1, we show an example of the proton-induced cascade emission.  The $p \gamma$ and $pp$ reactions lead to very high-energy gamma rays which eventually cascade down via pair creation, synchrotron and Compton emissions.  (Note that we include both IC scattering and Compton downscattering.)  The proton synchrotron component also contributes to the intrinsic very high-energy component but it is significantly attenuated and cascaded in the source.   
Then, as expected before, a flat and hard spectrum, which is different from the Band function, is typically obtained below the pair-creation cutoff.  Note that such a flat and hard spectrum is expected for larger dissipation radii as studied in detail by Asano et al. (2009a). 

\begin{figure}[tb]
\includegraphics[width=1.00\linewidth]{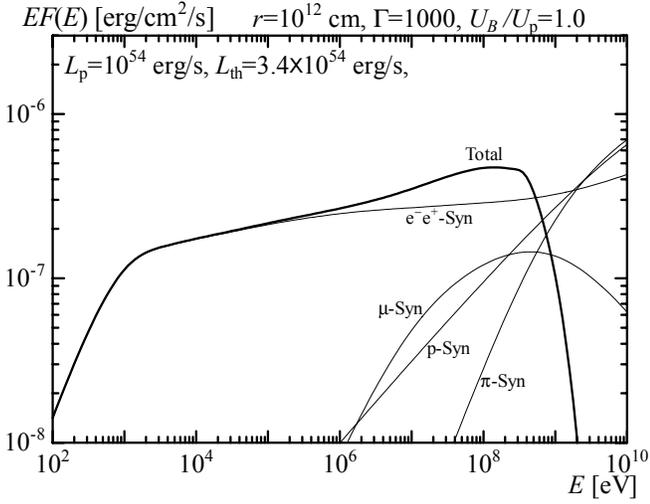}
\caption{Gamma-ray spectra of proton-induced cascade emission without stochastic re-acceleration. The source redshift is assumed to be $z=1$.  The strong magnetic field is assumed over the length scale of $l_{\rm tur}=0.17 l$ and $\eta=1$ is assumed.  The corresponding Thomson depth is $\tau_T=1$. }
\label{fig2}
\end{figure}

Now, we take into account effects of stochastic acceleration (or heating).  We give the energy gain/loss per collision, $\Delta \varepsilon/\varepsilon$, following the Gaussian probability function, according to Asano \& Terasawa (2009) that focused on the case of low-energy injection where the cascade does not play an important role.  Then, we set the stochastic acceleration (or heating) rate such that the synchrotron peak of $\gamma_{\rm typ}$ electrons correspond to $E^{\rm br}$.  
We also introduce sub-parameters, $\gamma_{\rm reacc}$ and $l_{\rm tur}$, but our main results are not sensitive to those additional parameters.  The former is related to how long stochastic acceleration (or heating) operates, and $\gamma_{\rm reacc} \ll \gamma_{\rm typ}$ is needed for the re-acceleration/heating to work well.  Since the calculations for $\gamma_{\rm reacc} \sim 1$ are too time-consuming, we take $\gamma_{\rm reacc} \sim 10-100$, which is enough for our purpose.  The latter parameter is physically motivated by the finite lifetime of the turbulent magnetic field, and we assume $l_{\rm tur}/l \sim 0.1-1$ in this work.  Thus, our ``one-zone'' code divides the shocked region into two parts: a strongly magnetized region of scale $l_{\rm tur}$ and the second, remaining part of scale $l-l_{\rm tur}$.  The photon density is assumed to be homogeneous over the entire scale $l$, and secondary particles from protons would be injected also homogeneously in this scale.  But, for simplicity, we treat only a fraction $l_{\rm tur}/l$ of the secondary particles injected in the strongly magnetized region.  Although IC emission by the particles injected in the non-disturbed region may appear in the GeV range~\citep{AT09}, we neglect such emission.  The absence of heating in this region may make such emission less important.  The parameter $l_{\rm tur}$ can affect high-energy IC spectra quantitatively, but the qualitative features of our results are not altered (see below).  The calculation time is set to $t_{\rm tur}=3 l_{\rm tur}/c$ for the secondary particles, while that for accelerated protons is $l/c$.  We do not consider stochastic acceleration of protons since more efficient shock acceleration is assumed.  We also remark that the energy ranges for pion-producing protons and electrons with $\sim \gamma_{\rm typ}$ are quite different.  Also, we neglect re-acceleration of pions and muons, though it might be potentially relevant (see appendix A).

In Figure~2, we show the case where stochastic re-acceleration (or heating) is included, for the same parameter set shown in Figure~1.  When the re-acceleration/heating is turned on, electrons and positrons can avoid their fast cooling, and the energy balance between acceleration/heating and cooling gives the typical energy of $\gamma_{\rm typ} \sim 100$.  Synchrotron emission forms another component in addition to the flat proton-induced cascade component, and its peak can be attributed to the observed peak energy.  The non-linear effect on the photomeson production enhances the proton-induced cascade component compared with the case in Figure~1.  As argued in Section~2.2, one sees that a hard electron distribution indeed leads to the low-energy photon index of $\alpha \sim 1$.  The high-energy synchrotron spectrum of re-accelerated/heated particles is expected to become steeper and steeper above the synchrotron peak.  But, as a result of the superposition of the slow heating component and the underlying cascade component, the high-energy photon index is effectively regarded as $\beta \sim 2.5$ at $\sim 1-10$~MeV energies.  As demonstrated in Figure~2, in the slow heating scenario with the electromagnetic cascade initiated by some injection at high energies, the resulting spectrum consists of an apparent broken power-law component due to the synchrotron emission from re-accelerated/heated particles and an extra hard component from the cascade emission (including IC emission by re-accelerated/heated particles),  which is different from the previous work~\cite{AT09} that instead obtained a simple power law with a high-energy cutoff.  Such a spectrum may also be responsible for observed \textit{Fermi} GRBs, e.g., GRB 090926B~\cite{Ack+11}.  Note that the gamma-ray luminosity in the MeV range comes from re-accelerated/heated leptons whose energies are supplied by the plasma/MHD turbulence, so that it can be larger than a fraction of the proton luminosity that is transferred to electromagnetic components via decay of mesons and muons.    
The spectral break at $\sim 10$ keV is determined by the parameter $\gamma_{\rm reacc}$, below which the stochastic acceleration (or heating) becomes ineffective.  This parameter is kept as a free one in this simulation so that the quantitative value of this low-energy break does not carry strong implications.  This break's appearance is interestingly similar to the observed one reported in some of {\it Fermi} GRBs. 

\begin{figure}[tb]
\includegraphics[width=1.00\linewidth]{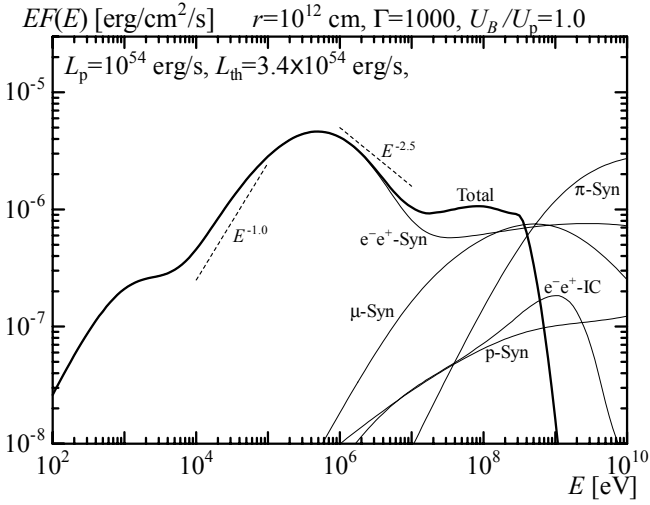}
\caption{Gamma-ray spectra of proton-induced cascade emission with stochastic acceleration (or heating). The source redshift is assumed to be $z=1$.  Used sub-parameters (that are not critical for the result) are $l_{\rm tur}/l=0.17$, $\gamma_{\rm reacc}=10$, and $\eta=1$.  The obtained gamma-ray luminosity is $L_\gamma \simeq 1.2 \times {10}^{53}~{\rm erg}~{\rm s}^{-1}$ and the corresponding Thomson depth is $\tau_T=1$.}
\label{fig2}
\end{figure}

In Figure~3, we show the fate of shock-accelerated high-energy protons that are injected with a simple power law.  The photomeson production efficiency is so high in this parameter set that almost all the protons are eventually depleted.  Note that UHECR production without their depletion is possible at larger dissipation radii~\cite{WB97,MINN08} but larger baryon loading would be required to achieve the same hadronic gamma-ray flux.  
One sees that high-energy mesons are mainly produced by the photomeson production, while lower-energy mesons come from the $pp$ reaction, and the result is in agreement with the analytical expectation described in Section~2.1.  Hence, when protons are accelerated up to sufficiently high energies, one expects that the photomeson production is typically dominant in cascaded gamma rays, though the signature of $pp$ reactions can be seen in TeV neutrinos for sufficiently small dissipation radii.  However, the $pp$ reaction can be more important when there is no proton acceleration~\citep[e.g.,][]{Bel10}, or the nonthermal proton spectral index is steeper than 2 or proton acceleration is inefficient~\cite{Mur08}.  Even in these cases, our model can work to reproduce the observed MeV emission.  Such a case is demonstrated in Figure~4, where we obtain a broken power-law spectrum that can be compatible with the Band function.  
The value of $\eta$ used here ($t_{{\rm acc},p}=\eta (E_p/\Gamma eB)$) seems much larger than conventional ones that are required for UHECR acceleration, although such values have been used to interpret observed blazar spectra~\citep[e.g.,][]{IT96}. 

\begin{figure}[tb]
\includegraphics[width=1.00\linewidth]{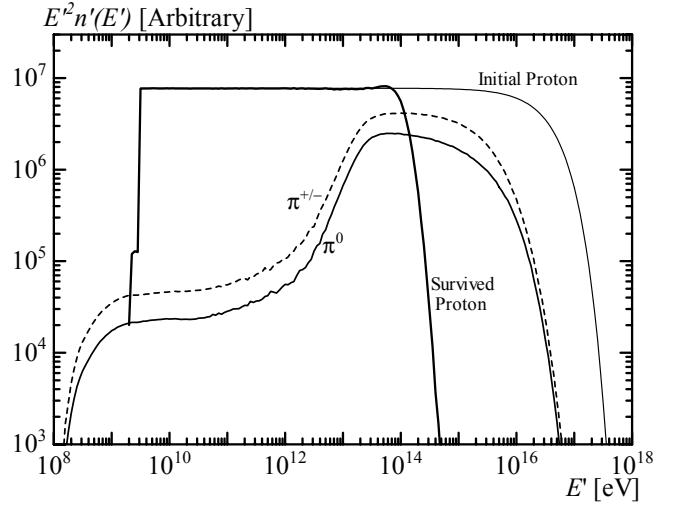}
\caption{Spectra of injected and survived protons in the outflow comoving frame for the calculation in Figure~2. The resulting pion spectra are also shown.}
\label{fig2}
\end{figure}
\begin{figure}[tb]
\includegraphics[width=1.00\linewidth]{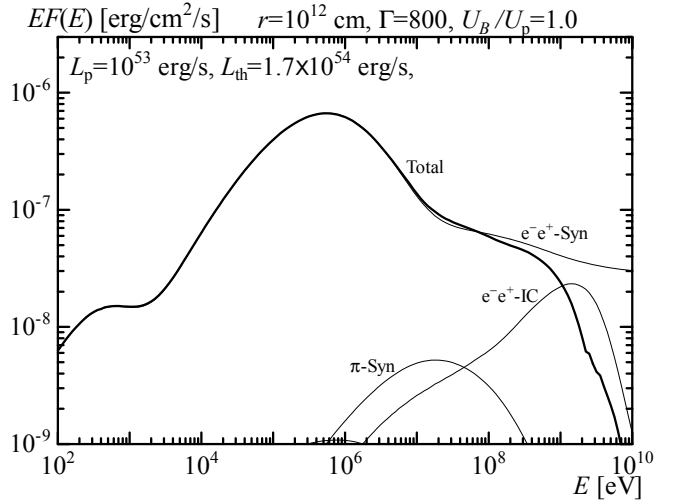}
\caption{Gamma-ray spectra for the case where proton acceleration is inefficient.  The source redshift is assumed to be $z=1$.  Used sub-parameters are $l_{\rm tur}/l=0.99$, $\gamma_{\rm reacc}=10$, and $\eta={10}^{4}$.  The obtained gamma-ray luminosity is $L_\gamma \simeq 1.6 \times {10}^{52}~{\rm erg}~{\rm s}^{-1}$ and the corresponding Thomson depth is $\tau_T=1$.}
\label{fig2}
\end{figure}

In Figure~5, we show the result for larger emission radii, taking into account effects of stochastic acceleration (or heating).  In this figure, a smaller magnetic field is assumed, so that IC emission from re-accelerated/heated leptons, which is expected around $5~{\rm GeV}~\gamma_{\rm typ,2}^2 (E^{\rm br}/500~\rm keV)$, become more prominent.  The effect of the IC emission is generally non-linear, and  the dependence on $l_{\rm tur}$ is demonstrated in Figure~6.  One can see that, for larger values of $l_{\rm tur}$, more particles are re-accelerated/heated and lose their energies via the IC emission.  As a result, more gamma rays are attenuated and cascaded, which enhance the extra hard component and also provide more seed photons for the photomeson prediction.  The different efficiency for IC emission results in the variety of the high-energy spectral index $\beta$, which is from $2.8$ to $\sim 2$ in Figure~6.  One also sees that $l_{\rm tur}$ cannot be too small to have sufficiently large gamma-ray luminosities in the MeV range, but the result is not so sensitive as long as $l_{\rm tur}$ is large enough.  Note that, the other sub-parameter, $\gamma_{\rm reacc}$, affects the flux ratio of the broken power-law component to the power-law-like cascade component rather than $l_{\rm tur}$, since more leptons can be injected to the re-acceleration/heating process for smaller $\gamma_{\rm reacc}$.  We do not show such other cases because of the onerous calculation time involved, but our qualitative results are not much altered by these sub-parameters as long as $\gamma_{\rm reacc} \gg \gamma_{\rm typ}$ and $l_{\rm tur}$ is large enough.  This is because the timescale for remaining above $\gamma_{\rm reacc}$ would be close to the diffusion timescale in energy space, $\sim \varepsilon^2/(4D_{\varepsilon \varepsilon})=t_{\rm sta}$.  Therefore, it would not be extremely large (${\mathcal E}_{\gamma \rm ex}/{\mathcal E}_{\gamma} \gtrsim 0.01$), unless the contribution from primary electrons become relevant.  Although we have considered relatively small emission radii, such a spectrum composed of an apparent broken-power law and an extra hard component is expected at larger radii too as long as high-energy injection and efficient cascades are achieved. 

\begin{figure}[tb]
\includegraphics[width=1.00\linewidth]{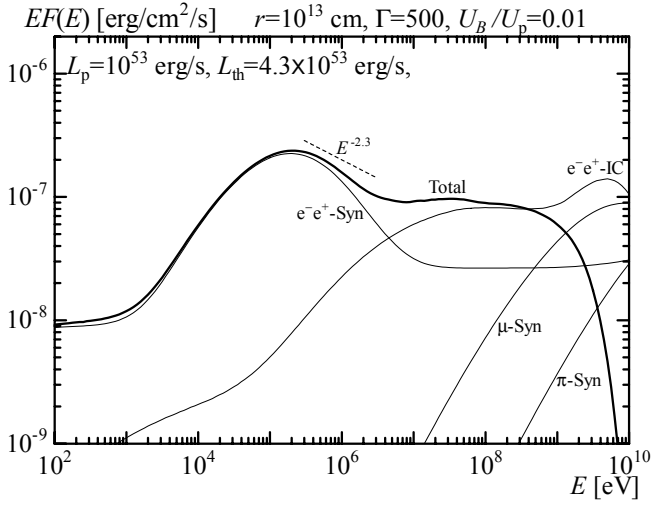}
\caption{Gamma-ray spectra of proton-induced cascade emission with stochastic acceleration (or heating). The source redshift is assumed to be $z=1$.  Used sub-parameters are $l_{\rm tur}/l=0.36$, $\gamma_{\rm reacc}=100$, and $\eta=1$.  The obtained gamma-ray luminosity is $L_\gamma \simeq 8.6 \times {10}^{51}~{\rm erg}~{\rm s}^{-1}$ and the corresponding Thomson depth is $\tau_T=0.1$.}
\label{fig2}
\end{figure}
\begin{figure}[tb]
\includegraphics[width=1.00\linewidth]{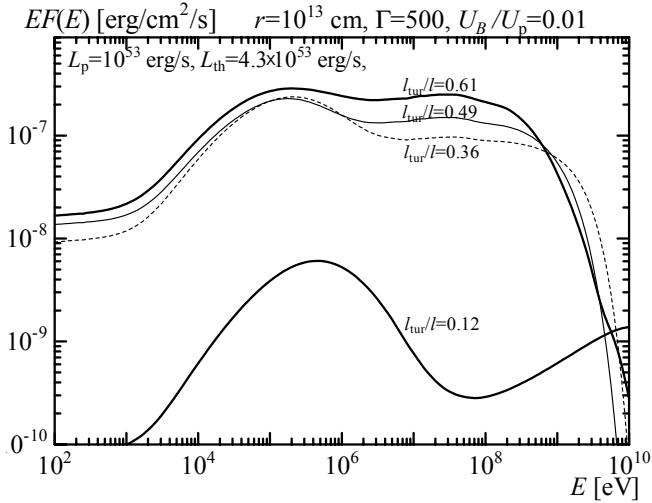}
\caption{Same as Figure~5, but dependence on the sub-parameter, $l_{\rm tur}$, is shown.}
\label{fig2}
\end{figure}

\section{Specific Implications of Hadronic Models for High-Energy Neutrino Observations}
The purpose of our paper is to investigate the role of stochastic acceleration (or wave heating), and we specifically consider hadronic processes as high-energy injection processes.  At the same time, it would be useful to discuss implications of hadronic models that can explain observed gamma rays.  One of the important tests of the hadronic models is the prediction of high-energy neutrinos.  The high-energy neutrino emission is a generic consequence of hadronic models which explain observed gamma-ray spectra through proton-induced cascade emission rather than via proton synchrotron emission~\citep[e.g.,][]{BHMO10}, whether the re-acceleration/heating occurs or not.  In the hadronic model explaining the extra hard component with the proton-induced cascade, one typically expects, ${\mathcal E}_{\nu} \sim {\mathcal E}_{\gamma \rm ex}$.  This relation between ${\mathcal E}_\nu$ and ${\mathcal E}_{\gamma \rm ex}$ is not much affected by the multi-pion production effect, even though estimates of $f_{\rm mes}$ and the required $\mathcal{E}_{\rm CR}$ can be affected.  However, we do not know typical values of ${\mathcal E}_{\gamma \rm ex}$ at present, and recent 
analyses indicate that the GeV component is less than 10~\% of the MeV one on average~\citep[e.g.,][]{BGNP11}.  

In Figure~7, we show neutrino spectra for three parameter sets shown in the previous section, when meson re-acceleration is irrelevant.  For normalization, the total gamma-ray energy is set to ${\mathcal E}_\gamma^{\rm iso}={10}^{53.5}$~erg and $z=1$ is assumed.  For the thick curve where $\tau_T=0.1$, PeV-EeV neutrinos coming from the photomeson production are mainly expected.  For the dashed curve where $\tau_T=1$, $pp$ neutrinos become important, as in the dissipative photospheric scenario~\cite{Mur08,WD09} and resulting in similar neutrino spectra, although our prediction comes from an independent motivation so that a larger baryon loading is typically needed to have a visible $\sim$~GeV component. 
For the dot-dashed curve, only $pp$ neutrinos are relevant because inefficient acceleration is assumed, and we do not expect $\sim$~PeV neutrinos because the proton maximum energy is not so high.  In any case, for one burst at $z \sim 1$, the expected number of muon events in IceCube is $\sim {10}^{-2}-{10}^{-1}$, which seems too small to detect.  Hence, bright bursts with $\mathcal{E}_\nu^{\rm iso} \gtrsim {10}^{54}$~erg or nearby bursts at $z \sim 0.1$ are typically necessary~\cite{DA03}, although they would be rare.  

\begin{figure}[tb]
\begin{center}
\includegraphics[width=0.80\linewidth]{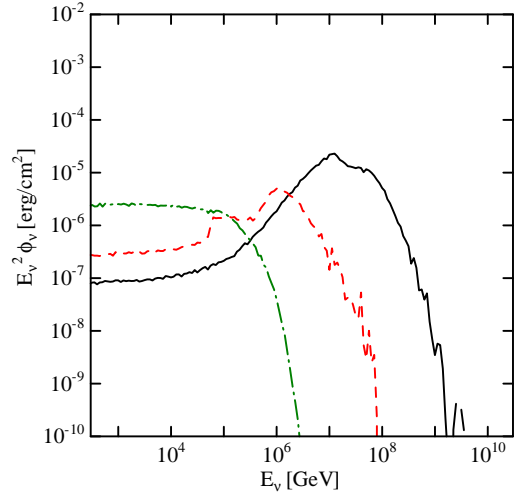}
\caption{The total neutrino fluences (for all flavors) from a GRB event at $z=1$. The dot-dashed, dashed and solid lines represent the fluences corresponding Figures~4, 3, and 5, respectively, in order of the relative importance of the $pp$ reaction.}
\label{fig2}
\end{center}
\end{figure}

On the other hand, the cumulative neutrino background is of interest, since time and space coincidences are expected to be of use for analyses of GRB prompt emission.  The cumulative neutrino background is calculated using the following formula, 
\begin{eqnarray}
\Phi_{\nu} = \frac{c}{4\pi H_{0}} \int _{0}^{z_{\rm max}} dz \, 
\frac{dN_{\nu}((1+z)E_{\nu})}{dE_{\nu}^{'}} \frac{R(z)}{\sqrt{\Omega_{\Lambda}+\Omega_{m}{(1+z)}^{3}}},
\end{eqnarray}
where $dN_\nu/d E_\nu^{'}$ is the neutrino spectrum in the cosmological rest frame and $R(z)$ is the GRB rate.  For demonstration, we use the GRB3 evolution but differences in evolution models do not make significant changes~\citep[see][and references therein]{Mur07}.  

An example of the cumulative neutrino background expected in the hadronic model is shown in Figure~8, using the parameters given in the caption of Figure~5.  
However, the appropriate normalization of the background flux is not so obvious, since ${\mathcal E}_{\gamma}$ comes from the turbulent energy rather than the cosmic-ray energy in the re-acceleration/heating scenario, where ${\mathcal E}_{\gamma \rm ex}/{\mathcal E}_{\gamma}$ can be small theoretically.  Indeed, ${\mathcal E}_{\gamma \rm ex}/{\mathcal E}_{\gamma}$ should typically be smaller than the value inferred in Figure~5~\citep[e.g.,][]{BGNP11}, which can be achieved by changing sub-parameters (c.f. Figure~4).  Therefore, on an \textit{ad hoc} basis, the background flux is rescaled using ${\mathcal E}_{\gamma \rm ex}$ instead of ${\mathcal E}_{\gamma}$.  Since ${\mathcal E}_{\gamma \rm ex}$ is highly uncertain at present, for demonstration purposes, we introduce the differential gamma-ray energy at 100~MeV, $\tilde{\mathcal E}_{\gamma \rm ex}$, and rescale the background flux assuming that it is a few~\% of the total gamma-ray energy.   The muon neutrino flux is analytically estimated to be
\begin{equation}
E_\nu^2 \Phi_\nu \sim {10}^{-9}~{\rm GeV}~{\rm cm}^{-2}~{\rm s}^{-1}~\tilde{\mathcal E}_{{\rm HE} \gamma {\rm ex},42} (f_z/3),
\end{equation}
where $f_z$ is the correction due to the redshift evolution, and the characteristic neutrino energy is generally model-dependent (cf. Figure~7).  One should also keep in mind the following limitations of the result.  (1) We have used the one-zone neutrino spectrum, but contributions from larger emission radii would be dominated at high energies~\citep[e.g.,][]{MN06a}.  (2) The background spectrum can be affected by individual characteristics of the bursts~\citep[e.g.,][]{BSHR06}, so that it would be valid for the purpose of estimating the flux level when typical parameters are chosen.  In our case, if the extra hard component at $\gtrsim 10$~MeV energies is less prominent for most of the bursts, the background flux should also be less.  Despite these caveats, the figure suggests that high-energy neutrino signals should be one of the important messengers of the hadronic model, and one may expect a few events per year in IceCube, provided the model assumptions are valid.  Just for comparison, predictions in other scenarios are also shown in Figure~8. 

\begin{figure}[tb]
\begin{center}
\includegraphics[width=0.80\linewidth]{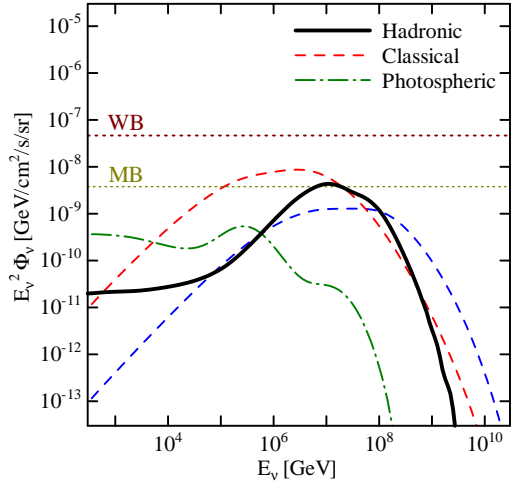}
\caption{The cumulative neutrino backgrounds (for all flavors) from GRBs.  Hadronic: calculated in this work for the demonstrative parameter set used in Figure~5, with rescaled normalization of $\tilde{\mathcal E}_{\rm HE\gamma \rm ex} R(0) = {10}^{42}~{\rm erg}~{\rm Mpc}^{-3}~{\rm yr}^{-1}$.  Classical: originally predicted by Waxman \& Bahcall (1997) and numerically calculated spectra (sets A and B) are taken from Murase \& Nagataki (2006a), but with normalization of $\tilde{\mathcal E}_{\rm HECR} R(0) = 5 \times {10}^{43}~{\rm erg}~{\rm Mpc}^{-3}~{\rm yr}^{-1}$.  Photospheric: calculated in Murase (2008), but with normalization of $\xi_{\rm CR} \equiv {\mathcal E}_{\rm CR}/{\mathcal E}_{\gamma}=1$ and $\mathcal{E}_{\gamma}^{\rm iso}={10}^{53.5}$~erg.  WB: the Waxman-Bahcall bound~\cite{WB99} shown as landmarks.  MB: the effective iron-survival bound~\cite{MB10} shown for comparison.  
}
\label{fig2}
\end{center}
\end{figure} 

Recently, the IceCube collaboration~\cite{Abb+11} has given upper limits on the neutrino background flux~\footnote{Note that time- and space-coincidence between neutrinos and gamma rays is assumed~\citep[but see, e.g.,][]{JP07}} (with a normalization procedure different from theoretical work by Waxman \& Bahcall 1997 and Murase \& Nagataki 2006).  Such limits are interesting in order to test the GRB-UHECR proton hypothesis for the prompt phase~\cite{WB97} as the photomeson production efficiency of $f_{p \gamma} \lesssim 1$ can be probed.  Generally speaking, the efficiency is model-dependent.  One typically expects ${\rm max} [1, f_{p \gamma}] \sim 0.1-1$ for emission radii of $r \sim {10}^{13}-{10}^{14.5}$~cm, while $f_{p \gamma} \sim {10}^{-2.5}-{10}^{-1}$ for $r \sim {10}^{14}-{10}^{15.5}$~cm~\citep[][]{MN06a}.  Especially in the neutron escape scenario where the neutrino flux is comparable to the escaping cosmic-ray flux~\citep[e.g.,][]{MPR00}, the limits are already stringent for steep proton spectra~\cite{AGH11}.  However, the detection of neutrinos would typically be difficult if the observed UHECRs are mainly heavy nuclei rather than protons~\cite{MINN08,MB10}.  Observations in the near future will give more stringent 
limits on $f_{p \gamma}$, based on the GRB-UHECR proton hypothesis.  
On the other hand, the hadronic model focused on in this work relies on the different motivation of explaining the observed GeV emission, where we do not have to explain UHECRs by GRBs.  Since it is uncertain how common and energetic the extra hard component at $\gtrsim 10$~MeV is, it is not so obvious to derive quantitative constraints on the model.  Nevertheless,  although the background flux seems lower than the current limits, our result implies that future neutrino observations would be crucial for the hadronic models provided that the extra hard component commonly carries $1-10$~\% of the MeV component.  Limits at differential energies would especially be useful. 

\section{Summary and Discussion}
In this paper we studied the role of stochastic acceleration (or wave heating) on the GRB prompt emission, in the presence of electromagnetic cascades.  As an example, we employ the hadronic model to initiate cascade processes.  We demonstrated that some of the current issues in the GRB prompt emission can be explained.  (1) The low-energy photon index is expected to be $\alpha \sim 1$ when electrons and positrons confined in the downstream region are accelerated stochastically or heated via turbulence, and the typical peak energy, $E^{\rm br}$, is stabilized by the balance between the stochastic acceleration (or heating) time and the cooling time.  
(2) It is possible to avoid the fast cooling problem because the charged leptons are slowly heated by turbulence.  When the leptons are injected via the hadronic processes, we may not have a potential issue on injection of primary electrons in the shock acceleration mechanism.  
(3) An extra hard component, $E F(E) \propto E^{0-0.5}$, which has often been observed above $1-100$~MeV, can be explained by cascade emission, such as in proton-induced cascades.  The existence of the extra component is an important feature of the cascade initiated by high-energy injection well above MeV energies.  

Although we have considered one-zone cases, dissipation will occur at various radii, which can affect the high-energy spectrum.  As shown in appendix B, in slow heating scenarios via stochastic acceleration, the high-energy photon spectrum can be harder than in the one-zone case because of the superposition of one-zone slow-heating spectra.  As a result, the broadband spectral shape can be a broken power law up to $\sim$~GeV energies.  Such multi-zone effects are even more important above $\gtrsim 0.1$~GeV.  Smaller $\gamma \gamma$ opacities at larger radii generally smear out the $\gamma \gamma$ pair-creation cutoff~\cite{Aoi+10}.  Note that, in our calculation method (where spectra are obtained via iteration), the radiative transfer effect in the emitting slab, which leads to $(1-\exp(-\tau_{\gamma \gamma}))/\tau_{\gamma \gamma}$ rather than the exponential cutoff, is not included.  Also, time-dependent and geometrical effects introduce additional complications, and the pair-creation break can be higher~\cite{GTS08}.  Hence, we expect that the extra component extends to higher energies, although details are beyond the scope of this work.   

One of the observational properties found by \textit{Fermi} is the delayed onset at GeV energies.  It is not so obvious how to explain this in our model.  However, there are several reasons that may lead to broader pulses at high energies.  Especially, multi-zone effects seem important.  As expected from the above, the cascade component as well as the slow heated component would be superposed by dissipation at various radii.  Also, gamma rays from outer radii would be more transparent, which could make a dominant contribution at GeV-TeV energies.  In addition, in the hadronic models, neutron beams from the inner radius make further gamma rays via neutron decay and photomeson production with low-energy photons produced by dissipation at the outer radius.  Leptons may also up-scatter the MeV emission from inner radii via the external IC process.   
In addition, in our model, emissions on timescales longer than $t_{\rm tur}$ and $t_{\rm dyn}$ are expected.  Even after the dynamical time, the residual energy of electrons would be released by the external IC process and synchrotron radiation in some residual magnetic field.  In the hadronic models, further hadronic reactions may also happen as the escaping cosmic rays diffuse in the residual magnetic field (although they are adiabatically cooled).  Hence, future refinements based on multi-zone calculations seem necessary.   

All hadronic models have typically several caveats or open issues.  They often lead to the requirement of a high baryon loading, which may be problematic, although this is motivated mainly by the GRB-UHECR hypothesis.  Another potential issue is the radiative efficiency.  In our cases, we typically obtain the reasonable radiative efficiency, $L_\gamma/L_{\rm th} \sim 0.1$, although the apparent radiative efficiency can be higher due to lower $\gamma_{\rm reacc}$, release of the cosmic rays, residual emission and subsequent internal collisions.  Despite these concerns, there are interesting features.  Importantly, the hadronic model seems testable by upcoming multi-messenger observations.  The high-energy neutrino signal is one of the most important signatures, since the neutrino luminosity is expected to be comparable to the extra hard component.  If this extra component is ubiquitous and it carries $\sim 1-10$~\% of the total gamma-ray energy, IceCube would detect this signal in multi-year observations.  It is also important to check whether the extra hard component is common among all GRBs.  
As of now, its detectability with \textit{Fermi} is limited and the current analyses suggest that the GeV emission has less than $\sim 10$~\% of the MeV emission.  Therefore, in order to unravel the properties of the high-energy emission, deeper observations via detections with Cherenkov detectors such as MAGIC and VERITAS should be important, even though detections from distant GRBs become difficult due to the attenuation by the extragalactic background light.  Although very-high-energy photons from GRBs have not been firmly detected so far, the future CTA~\cite{CTA10} and HAWC are anticipated to change this situation.   
 
We note that, although we have focused on the hadronic model as an example of high-energy injection, our results can be applied to other models, e.g., such as magnetic  dissipation.  In the slow heating scenario resulting from stochastic acceleration (or heating), it seems common to have hard low-energy photon indices, $\alpha \sim 1$ and $\beta >2$.  
This feature itself would not be changed by details of the high-energy injection, as long 
as slow heating can operate and the MeV peak comes from balance between acceleration/heating and cooling.  
Hence, as for the origin of stochastically accelerated (or heated) particles, one may think of other 
high-energy injection processes and subsequent cascades.  If magnetic dissipation such as 
reconnection accelerates electrons up to very high energies and subsequent cascades are 
developed, similar spectra would be obtained.  

\begin{acknowledgements} 
K.M. is greatly thankful to Tatsumi Koi for technical support to use Geant 4, and a part of calculations was carried out on Altix3700 BX2 at YITP in Kyoto University.  We thank Uri Keshet, Soebur Razzaque and Kenji Toma for useful suggestions and discussions.  K.M. acknowledges financial support by a Grant-in-Aid from JSPS, from CCAPP and from PSU.  K.M. also thanks the hospitality of TIT and PSU, during a part of this project.  This study is partially supported by Grants-in-Aid for Scientific Research No.~22740117 from the Ministry of Education, Culture, Sports, Science and Technology (MEXT) of Japan (K.A.).  P.M. acknowledges partial support from NASA NNX08AL40G and NSF PHY-0757155. 
\end{acknowledgements}

\appendix
\section{Stochastic Re-acceleration of Mesons and Muons}
Here, we consider the possible effect of re-accelerated mesons on high-energy neutrino spectra. 
First, for $E_{\nu}^{\rm br}< E_{\nu}^{s}$, the classical neutrino spectrum produced by shock accelerated protons via the photomeson production is written as~\cite{WB99,RM98,Wax11,Mur07} 
\begin{equation}
E_{\nu}^2 \phi_{\nu} \propto
\left\{ \begin{array}{ll}
{(E_{\nu}/E_{\nu}^{\rm br})}^{\beta-1} 
& \mbox{($E_{\nu} \leq E_{\nu}^{\rm br}$)}
\\
{(E_{\nu}/E_{\nu}^{\rm br})}^{\alpha-1} 
& \mbox{($E_{\nu}^{\rm br} < E_{\nu} \leq E_{\nu}^{s}$)}
\\
{(E_{\nu}^s/E_{\nu}^{\rm br})}^{\alpha-1}{(E_{\nu}/E_{\nu}^{s})}^{\alpha-3}
& \mbox{($E_{\nu}^{s}< E_{\nu}$)}
\end{array} \right.
\end{equation}
Here $E_{\nu}^{\rm br}$ is the low-energy break coming from $E_{\nu}^{\rm br} \approx 0.05 E_p^{\rm br}$ or the characteristic energy where $f_{p \gamma} (20 E_\nu)=1$~\cite{Asa05,MN06a}. 
Also, we have only assumed the pion synchrotron loss as the important cooling process, and $E_{\nu}^{s} \approx 0.25 E_{\pi}^s$ is the high-energy break coming from the pion synchrotron loss, though other processes such as adiabatic cooling could also be relevant.  
From $t_{\pi, \rm syn} = t_{\pi} \equiv \gamma_{\pi} \tau_\pi$ (where $\tau_\pi$ is the lifetime of charged pions), the high-energy break energy of pions is estimated to be
\begin{eqnarray}
E_{\pi}^s \approx \Gamma \sqrt{\frac{6 \pi {(m_{\pi} c^2)}^5}{\sigma_T m_e^2 c^5 B^2 \tau_{\pi}}} 
\simeq 190~{\rm PeV}~\epsilon_{B,-1}^{-1/2} r_{13} L_{\rm th, 53.5}^{-1/2} \Gamma_{2.7}^2.
\end{eqnarray}
Note that one can make similar estimates for other particles such as kaon and muons, changing proper mass and lifetime.  

If there is re-acceleration by turbulence, $t_{\pi, \rm sta}=t_{\pi, \rm syn}$ gives the maximum energy of accelerated pions unless the energy fraction carried by mesons exceeds $\epsilon_\pi$, similarly to the case of electrons.  When one uses equation~(9), the typical energy of pions is estimated to be
\begin{equation}
E_{\pi}^{\rm typ} \approx 
\Gamma {\left[ \frac{6 \pi m_{\pi}^4 c^4 {(e B)}^{2-q}}{\sigma_T m_e^2 B^2 l_{\rm tur}^{q-1} \eta_{\rm sta}} \right]}^{\frac{1}{3-q}}.
\end{equation}
The situation depends on details of the plasma/MHD turbulence at relevant scales.  As numerical examples, we have $E_{\pi}^{\rm typ} \simeq 120~{\rm PeV}~\epsilon_{B,-1}^{-3/4} r_{13}^{3/2} l_{\rm tur,10}^{-2/3} L_{\rm th,53.5}^{-3/4} \Gamma_{2.7}^{5/2} \eta_{\rm sta}^{5/6}$ for $q=1.8$, and $E_{\pi}^{\rm typ} \simeq 880~{\rm PeV}~\epsilon_{B,-1}^{-1/2} r_{13} l_{\rm tur,10}^{-1/3} L_{\rm th,53.5}^{-1/2} \Gamma_{2.7}^{2} \eta_{\rm sta}^{2/3}$ for $q=1.5$.  However, the re-acceleration of charged mesons is possible only when particles are injected to the re-acceleration process before their decay. The critical energy is determined by $t_{\pi} = t_{\pi,\rm sta}$, and we find
\begin{equation}
E_{\pi}^{\rm cr} \approx \Gamma {\left( \frac{\tau_\pi {(eB)}^{2-q} l_{\rm tur}^{1-q}}{m_\pi c \eta_{\rm sta}} \right)}^{\frac{1}{1-q}}.
\end{equation}
For $1<q<2$, we expect two cases, $E_{\pi}^{\rm cr} < E_{\pi}^s < E_{\pi}^{\rm typ}$ and $E_{\pi}^{\rm typ} < E_{\pi}^s < E_{\pi}^{\rm cr}$, and re-acceleration of mesons is relevant only in the former case at energies above $E_{\pi}^{\rm cr}$.  For $q<1$, two cases, $E_{\pi}^s < E_{\pi}^{\rm typ} < E_{\pi}^{\rm cr}$ and  $E_{\pi}^{\rm cr} < E_{\pi}^{\rm typ} < E_{\pi}^s$, are possible, and mesons with energies below $E_{\pi}^{\rm cr}$ are re-accelerated in the former case.  Usually such re-acceleration will not be important at large dissipation radii especially above the photospheric radius, since time scales such as $t_{\rm sta}$, $t_{\rm cool}$ and $t_{\rm dyn}$ become longer for larger radii whereas $t_\pi$ is unaffected.    
If the meson re-acceleration happens, mesons are re-distributed so that the resulting meson spectrum will be altered.  As long as particle escape is irrelevant, the neutrino spectrum can be hard, and one may expect 
\begin{equation}
E_{\nu}^2 \phi_\nu \propto E^{3-q} \,\, \mbox{($E_{\nu} \leq E_\nu^{\rm typ}$)}.
\end{equation}
The typical neutrino energy can be very high energies of $\sim 100$~PeV, so that detectability of such very high-energy neutrinos may be enhanced by re-acceleration. 

\section{Superposition of Slow Heating Spectra}
In the slow heating scenario with stochastic acceleration (or heating), whether the origin of relativistic electrons (and positrons) is leptonic or hadronic, one may expect a synchrotron component with a hard spectrum, 
\begin{equation}
E F(E) \propto E^{2-\alpha} \,\, \mbox{($E \leq E^{\rm br}$)},
\end{equation}
where $\alpha \sim q/2$ if the MHD turbulence is responsible for stochastic acceleration. Above $E^{\rm br}$, the spectrum would typically have a cutoff feature or significant steepening ($\beta \gg 2$). 

So far, we have assumed the one-zone case, where dissipation occurs at the specific radius, $r$. But, realistically, dissipation continues at further radii, and we discuss its effect on the high-energy spectrum. For simplicity, let us here assume that $L_{\gamma}, L_B \propto L_{\rm diss} \propto r^{-s}$ ($s>0$).  For a given $\Gamma \propto r^{-x}$, assuming $l_{\rm tur} \propto l \propto r/\Gamma$, one obtains 
\begin{equation}
t_{\rm sta} \propto B^{q-2} \gamma^{2-q} l_{\rm tur}^{q-1} \propto r^{q-1-(q/2-1)(s+2)+x(2q-3)} \gamma^{2-q}. 
\end{equation} 
On the other hand, the (synchrotron/IC) cooling timescale has 
\begin{equation}
t_{\rm cool} \propto r^{2} \Gamma^2 L_{\rm diss}^{-1} \gamma^{-1} \propto r^{s+2-2x} \gamma^{-1}.
\end{equation}
Then, for the synchrotron break energy, we have 
\begin{equation}
E^{\rm br} \propto \Gamma \gamma_{\rm typ}^2 B \propto r^{(sq+2q-4)/(3-q)+1-s/2+2x(1-2q)/(3-q)}. 
\end{equation}
As a result, if the one-zone high-energy photon spectrum above $E^b$ is steep enough, the multi-zone high-energy photon spectrum may have a harder one,  
\begin{equation}
E F(E) \propto E^{-\frac{2s(3-q)}{2(sq+2q-4)-(s-2)(3-q)+4x(1-2q)}}.
\end{equation}
Especially if $q \sim 2$ and $x \sim 0$, one has $E F(E) \propto E^{-s/(1+1.5s)}$ which may be responsible for the high-energy index. In this case, higher-energy gamma rays are emitted mainly from larger radii, so that those timescales can in principle be longer. 
Note that the above discussion can be applied to not only the shock dissipation models but also the magnetic dissipation models, and a broken power-law spectrum can be expected in either case.


\begin{thebibliography}{}
\bibitem[Abbasi et al. 2011]{Abb+11}
Abbasi, R., et al. 2011, Phys. Rev. Lett., 106, 141101
\bibitem[Abdo et al. 2009a]{Abd+09a}
Abdo, A.~A., et al. 2009a, Science, 323, 1688
\bibitem[Abdo et al. 2009b]{Abd+09b}
Abdo, A.~A., et al. 2009b, 706, L138
\bibitem[Abdo et al. 2009c]{Abd+09c}
Abdo, A.~A., et al. 2009c, ApJ, 707, 580
\bibitem[Abdo et al. 2010]{Abd+10}
Abdo, A.~A., et al. 2010, ApJ, 712, 558
\bibitem[Ackermann et al. 2010]{Ack+10}
Ackermann, M., et al. 2010, ApJ, 712, 558
\bibitem[Ackermann et al. 2011]{Ack+11}
Ackermann, M., et al. 2011, ApJ, 729, 114
\bibitem[Ahlers et al. 2011]{AGH11}
Ahlers, M., Gonzalez-Garcia, M.~C., \& Halzen, F. 2011, arXiv:1103.3421
\bibitem[Agostinelli et al. 2003]{Ago+03}
Agostinelli, S., et al. 2003, Nucl. Instrum. Methods Phys. Res., Sect. A, 506, 250
\bibitem[Alfv\'en 1947]{Alf47}
Alfv\'en, H. 1947, MNRAS, 107, 211
\bibitem[Aoi et al. 2010]{Aoi+10}
Aoi, J., Murase, K., Takahashi, K., Ioka, K., \& Nagataki, S. 2010, ApJ, 722, 440
\bibitem[Asano 2005]{Asa05}
Asano, K. 2005, ApJ, 623, 967
\bibitem[Asano \& Inoue 2007]{AI07}
Asano, K., \& Inoue, S. 2007, ApJ, 655, 762
\bibitem[Asano \& Nagataki 2006]{AN06}
Asano, K., \& Nagataki, S. 2006, ApJ, 640, L9
\bibitem[Asano \& Takahara 2003]{AT03}
Asano, K., \& Takahara, T. 2003, PASJ, 55, 433
\bibitem[Asano \& Terasawa 2009]{AT09}
Asano, K., \& Terasawa, T. 2009, ApJ, 705, 1714
\bibitem[Asano et al. 2009a]{AIM09}
Asano, K., Inoue, S., \& M\'esz\'aros, P. 2009a, ApJ, 699, 953
\bibitem[Asano et al. 2009b]{AGM09}
Asano, K., Guiriec, S., \& M\'esz\'aros, P. 2009b, ApJ, 705, L191

\bibitem[Baerwald et al. 2011]{BHW11}
Baerwald, P., H\"ummer, S., \& Winter, W. 2011, Phys. Rev. D, 83, 067303
\bibitem[Baring 2006]{Bar06}
Baring, M.~G. 2006, ApJ, 650, 1004
\bibitem[Becker et al. 2006]{BSHR06}
Becker, J.~K., Stamatikos, M., Halzen, F., Rhode, \& W. 2006, Astropart. Phys., 25, 118
\bibitem[Becker et al. 2010]{BHMO10}
Becker, J.~K., Halzen, F., Murchadha, A.~O., \& Olivo, M. 2010, ApJ, 721, 1891
\bibitem[Bednarz \& Ostrowski 1996]{BO96}
Bednarz, J., \& Ostrowski, M. 1996, MNRAS, 283, 447
\bibitem[Beloborodov 2000]{Bel00}
Beloborodov, A. M. 2000, ApJ, 539, L25
\bibitem[Beloborodov 2010]{Bel10}
Beloborodov, A.~M. 2010, MNRAS, 407, 1033
\bibitem[Beniamini et al. 2011]{BGNP11}
Beniamini, P., Guetta, D., Nakar, E., \& Piran, T. 2011, arXiv:1103.0745
\bibitem[Beresnyak et al. 2011]{BYL11}
Beresnyak, A., Yan, H., \& Lazarian, A. 2011, ApJ, 728, 60
\bibitem[Blumenthal \& Gould 1970]{BG70}
Blumenthal, G.~R., \& Gould, R.~J. 1970, Rev. Mod. Phys., 42, 237
\bibitem[Brunetti \& Lazarian 2007]{BL07}
Brunetti, G., \& Lazarian, A. 2007, MNRAS, 378, 245
\bibitem[Bykov \& M\'esz\'aros 1996]{BM96}
Bykov, A.~M., \& M\'esz\'aros, P. ApJ, 461, L37

\bibitem[Chang et al. 2008]{CSA08}
Chang, P., Spitkovsky, A., \& Arons, J. 2008, ApJ, 674, 378
\bibitem[Cho et al. 2003]{Cho+03}
Cho, J., Lazarian, A., \& Vishniac, E. T. 2003, in Turbulence and Magnetic Fields in Astrophysics, ed. E. Falgarone \& T. Passot (Berlin: Springer), 56
\bibitem[Chodorowski et al. 1992]{CZS92}
Chodorowski, M.~J., A. Zdziarski, A.~A., \& Sikora, S.~R. 1992 ApJ, 400, 181
\bibitem[Corsi et al. 2010]{CGP10}
Corsi, A., Guetta, D., \& Piro, L. 2010, ApJ, 499, L31
\bibitem[CTA Consortium 2010]{CTA10}
CTA Consortium 2010, arXiv:1008.3703

\bibitem[Daigne et al. 2011]{DBD11}
Daigne, F., Bosnjak, Z., \& Dubus, G. 2011, A\&A, 526, 110
\bibitem[Derishev et al. 2001]{DKK01}
Derishev, E.~V., Kocharovsky, V.~V., \& Kocharovsky, VI.~V. 2001, A\&A, 372, 1071
\bibitem[Dermer \& Atoyan 2003]{DA03}
Dermer, C.~D., \& Atoyan, A. 2003, Phys. Rev. Lett., 91, 071102
\bibitem[Dermer \& Atoyan 2006]{DA06}
Dermer, C.~D., \& Atoyan, A. 2006, New. J. Phys., 8, 122 
\bibitem[Dermer et al. 1996]{DML96}
Dermer, C.~D., Miller, J.~A., \& Li, H. 1996, ApJ, 456, 106

\bibitem[Eichler \& Waxman 2005]{EW05}
Eichler, D., \& Waxman, E. 2005, ApJ, 627, 861

\bibitem[Fossum \& Carlsson 2005]{FC05}
Fossum, A., \& Carlsson, M. 2005, Nature, 435, 919

\bibitem[Ghisellini \& Celotti 1999]{GC99}
Ghisellini, G., \& Celotti, A. 1999, ApJ, 511, L93
\bibitem[Ghisellini et al. 2010]{Ghi+10}
Ghisellini, G., et al. 2010, MNRAS, 403, 926
\bibitem[Granot 2010]{Gra10}
Granot, J. 2010, arXiv:1003.2452
\bibitem[Granot et al. 2008]{GTS08}
Granot, J., Cohen-Tanugi, J., do Couto e Silva, E. 2008, ApJ, 677, 92
\bibitem[Gupta \& Zhang 2008]{GZ08}
Gupta, N., \& Zhang, B. 2008, MNRAS, 384, L11

\bibitem[He et al. 2011]{He+11}
He, H.~N., et al. 2011, ApJ, 733, 22
\bibitem[Hurley et al. 1994]{Hur+94}
Hurley, K., et al. 1994, Nature, 372, 652

\bibitem[Ioka 2010]{Iok10}
Ioka, K. 2010, Prog. Theo. Phys., 124, 667 
\bibitem[Ioka et al. 2006]{ITYN06}
Ioka, K., Toma, K., Yamazaki, R., \& Nakamura, T. 2006, A\&A, 458, 7
\bibitem[Ioka et al. 2007]{Iok+07}
Ioka, K., Murase, K., Toma, K., Nagataki, S., \& Nakamura, T. 2007, ApJ, 670, L77
\bibitem[Inoue \& Takahara 1996]{IT96}
Inoue, S., \& Takahara, F. 1996, ApJ, 463, 555
\bibitem[Inoue et al. 2011]{IAI11}
Inoue, T., Asano, K., \& Ioka, K. 2011, ApJ, 734, 77

\bibitem[Jacob \& Piran 2007]{JP07}
Jacob, U., \& Piran, T. 2007, Nature Phys., 3, 87 
\bibitem[Jess et al. 2009]{Jes+09}
Jess, D.~B., et al. 2009, Science, 323, 1582

\bibitem[Katz et al. 2007]{KKW07}
Katz, B., Keshet, U., \& Waxman, E. 2007, ApJ, 655, 375
\bibitem[Kato 2005]{Kat05}
Kato, T.~N. 2005, Phys. of Plasmas, 12, 080705
\bibitem[Keshet et al. 2009]{Kes+09}
Keshet, U., Katz, B., Spitkovsky, A., \& Waxman, E. 2009, ApJ, 693, L127

\bibitem[Kumar \& Barniol Duran 2010]{KD10}
Kumar, P., \& Barniol Duran, R. 2010, MNRAS, 409, 226

\bibitem[Lithwick \& Sari 2001]{LS01}
Lithwick, Y., \& Sari, R. 2001, ApJ, 555, 540
\bibitem[Liu et al. 2006]{LMPF06}
Liu, S., Melia, F., Petrosian, V., \& Fatuzzo, M. 2006, ApJ, 647, 1099
\bibitem[Lyutikov 2006]{Lyu06}
Lyutikov, M. 2006, New. J. of Phys., 8, 119

\bibitem[Mannheim et al. 2000]{MPR00}
Mannheim, K., Protheroe, R.~J., \& Rachen, J.~P. 2000, Phys. Rev. D 63, 023003
\bibitem[McKinney \& Uzdensky 2011]{MU11}
McKinney, J. C., \& Uzdensky, D. A. 2011, arXiv:1011.1904
\bibitem[Medvedev 2000a]{Med00}
Medvedev, M.~V. 2000a, ApJ, 540, 704
\bibitem[Medvedev 2000b]{Med00b}
Medvedev, M.~V. 2000b, ApJ, 541, 811
\bibitem[Medvedev \& Spitkovsky 2009]{MS09}
Medvedev, M.~V., \& Spitkovsky, A. 2009, ApJ, 700, 956
\bibitem[M\'esz\'aros 2006]{Mes06}
M\'esz\'aros, P. 2006, Rep. Prog. Phys., 69, 2259
\bibitem[M\'esz\'aros \& Rees 2000]{MR00}
M\'esz\'aros, \& P., Rees, M. J. 2000, ApJ, 530, 292
\bibitem[Miller \& Ramaty 1989]{MR89}
Miller, J.~A., \& Ramaty, R. 1989, ApJ, 344, 973
\bibitem[Mizuno et al. 2011]{Miz+11}
Mizuno, Y., et al. 2011, ApJ, 726, 62
\bibitem[Murase 2007]{Mur07} 
Murase, K. 2007, Phys. Rev. D, 76, 123001
\bibitem[Murase 2008]{Mur08} 
Murase, K. 2008, Phys. Rev. D, 78, 101302(R)
\bibitem[Murase 2009]{Mur09} 
Murase, K. 2009, Phys. Rev. Lett., 103, 081102
\bibitem[Murase \& Beacom 2010]{MB10}
Murase, K., \& Beacom, J.~F. 2010, Phys. Rev. D, 81, 123001
\bibitem[Murase \& Ioka 2008]{MI08} 
Murase, K., \& Ioka, K. 2008, ApJ, 676, 1123
\bibitem[Murase \& Nagataki 2006a]{MN06a} 
Murase, K., \& Nagataki, S. 2006a, Phys. Rev. D, 73, 063002
\bibitem[Murase \& Nagataki 2006b]{MN06b} 
Murase, K., \& Nagataki, S. 2006b, Phys. Rev. Lett., 97, 051101
\bibitem[Murase et al. 2008]{MINN08} 
Murase, K., Ioka, K., Nagataki, S., \& Nakamura, T. 2008, Phys. Rev. D, 78, 023005

\bibitem[Pe'er et al. 2006]{PMR06}
Pe'er, A., M\'esz\'aros, P., \& Rees, M.~J. 2006, ApJ, 642, 995
\bibitem[Pe'er \& Zhang 2006]{PZ06}
Pe'er, A., \& Zhang, B. 2006, ApJ, 653, 454 
\bibitem[Pe'er et al. 2011]{Pee+11}
Pe'er, A., et al. 2011, arXiv:1007.2228
\bibitem[Petrosian \& Liu 2004]{PL04}
Petrosian, V., \& Liu, S. 2004, ApJ, 610, 550 

\bibitem[Rachen \& M\'esz\'aros 1998]{RM98}
Rachen, J.~P., \& M\'esz\'aros, P. 1998, Phys. Rev. D, 58, 123005
\bibitem[Razzaque 2011]{Raz11}
Razzaque, S. 2011, ApJ, 724, L109
\bibitem[Razzaque et al. 2010]{RDF10}
Razzaque, S., Dermer, C.~D., Finke, J.~D. 2010, OAJ, 3, 150
\bibitem[Rees \& M\'esz\'aros 1994]{RM94}
Rees, M.~J., \& M\'esz\'aros, P. 1994, ApJ, 430, L93

\bibitem[Sironi \& Spitkovsky 2011]{SS11}
Sironi, L., \& Spitkovsky, A. 2011, 726, 75
\bibitem[Spitkovsky 2008]{Spi08}
Spitkovsky, A. 2008, ApJ, 673, L39
\bibitem[Spruit et al. 2001]{SDD01}
Spruit, H.~C., Daigne, F., \& Drenkhahn, G. 2001, A\&A, 369, 694
\bibitem[Suzuki \& Inutsuka 2005]{SI05}
Suzuki, T.~K., \& Inutsuka, S. 2005, ApJ, 632, L49
\bibitem[Svensson 1987]{Sve87}
Svensson, R. 1987, MNRAS, 227, 403

\bibitem[Thompson 1994]{Tho94}
Thompson, C. 1994, MNRAS, 270, 480
\bibitem[Totani 1998]{Tot98}
Totani, T. 1998, ApJ, 509, L81 
\bibitem[Toma et al. 2011]{TWM11}
Toma, K., Wu, X.~F., \& M\'esz\'aros, P. 2011, arXiv:1002.2634
\bibitem[Tomczyk et al. 2007]{Tom+07}
Tomczyk, S., et al. 2007, Science, 317, 1192

\bibitem[Vietri 1995]{Vie95}
Vietri, M. 1995, ApJ, 453, 883
\bibitem[Vietri 1997]{Vie97}
Vietri, M. 1997, Phys. Rev. Lett., 78, 4328
\bibitem[Vurm et al. 2011]{VBP11}
Vurm, I., Beloborodov, A.~M., \& Poutanen, J. 2011, arXiv:1104.0394
\bibitem[Vurm \& Poutanen 2009]{VP09}
Vurm, I., \& Poutanen, J. 2009, ApJ, 698, 293

\bibitem[Wang \& Dai 2009]{WD09}
Wang, X.~Y., \& Dai, Z.~G. 2009, ApJ, 691, L67
\bibitem[Wang et al. 2009]{WLDM09}
Wang, X.~Y., Li, Z., Dai, Z.~G., \& M\'esz\'aros, P. 2009, ApJ, 698, L98
\bibitem[Waxman 1995]{Wax95}
Waxman, E. 1995, Phys. Rev. Lett., 75, 386
\bibitem[Waxman 2010]{Wax10}
Waxman, E. 2010, arXiv:1010.5007
\bibitem[Waxman 2011]{Wax11}
Waxman, E. 2011, arXiv:1101.1155
\bibitem[Waxman \& Bahcall 1997]{WB97}
Waxman, E., \& Bahcall, J. 1997, Phys. Rev. Lett., 78, 2292
\bibitem[Waxman \& Bahcall 1999]{WB99}
Waxman, E., \& Bahcall, J. 1997, Phys. Rev. D, 59, 023002

\bibitem[Yonetoku et al. 2004]{Yon+04}
Yonetoku, D., et al. 2004, ApJ, 609, 935

\bibitem[Zhang 2007]{Zha07}
Zhang, B. 2007, ChJAA, 7, 1
\bibitem[Zhang \& Yan 2011]{ZY11}
Zhang, B., \& Yan, H. 2011, ApJ, 726, 90
\bibitem[Zhang et al. 2007]{Zha+07}
Zhang, B., et al. 2007, ApJ, 655, 989
\bibitem[Zhang et al. 2009]{ZMW09}
Zhang, W., MacFayden, A., \& Wang, P. 2009, ApJ, 692, L40
\end{thebibliography}
\end{document}